\documentclass[twocolumn,amsmath,amssymb,pra,showpacs]{revtex4}

\usepackage{theorem}
\usepackage{enumerate}
\usepackage{amsmath}
\usepackage{amssymb}
\usepackage{graphicx}
\usepackage{epstopdf}
\usepackage{color}
\usepackage{euscript}
\usepackage{float}

\theorembodyfont{\rmfamily}

\newcommand{\lv}{\left \vert}
\newcommand{\rv}{\right \vert}

\newcommand{\ketbra}[2]{\lv #1 \rangle \langle #2 \rv}

\newcommand{\tr}{\mathrm{tr}}

\begin{document}

\title{
Thermal states of random quantum many-body systems
}%

\author{Yoshifumi Nakata, and Tobias J. Osborne}
\affiliation{Institute for Theoretical Physics, Leibniz University Hannover, Appelstrasse 2, 30167 Hannover, Germany}

\begin{abstract}
We study a distribution of thermal states given by random Hamiltonians with a local structure. We show that the ensemble of thermal states monotonically approaches the unitarily invariant ensemble with decreasing temperature if all particles interact according to a single random interaction and achieves a state $t$-design at temperature $O(1/\log(t))$. For the system where the random interactions are local, we show that the ensemble achieves a state $1$-design. We then provide numerical evidence indicating that the ensemble undergoes a phase transition at finite temperature.
\end{abstract}

\date{\today}


\maketitle

\section{Introduction}
In quantum many-body systems, the number of degrees of freedom increases exponentially with the number of particles.
This leads to difficulties for analysing their physics. 
One way to circumvent this difficulty is to assume random interactions, which are understood to be caused by inevitable impurities and disorder present in the physical system,
and study typical properties of random many-body Hamiltonians.
This idea is developed in random matrix theory
and provides a successful description of the complicated physics of heavy atoms, quantum chromodynamics, mesoscopic systems, quantum gravity, and quantum chaotic systems (see e.g. Ref.~\cite{M1990}).
A study of random Hamiltonians has recently been extended to quantum spin systems on a lattice~\cite{BS1970,BS1971,BF1971,BF1971-2,HMH2004,HMH2005,KLW2014,KLW2014-2}, where Hamiltonians contain only local interactions and respect the local structure of the system.
Such random {\it local} Hamiltonians were shown in Refs.~\cite{KLW2014,KLW2014-2} to have a distribution of eigenvalues different from that of random Hamiltonians without local structure, which we call random {\it global} Hamiltonians, implying that random local models are significantly distinct from global ones.

The idea of randomisation was applied to a study of the typical properties of quantum states
using the unitarily invariant ensemble of states, often called {\it random states}.
It has been pointed out that random states
play a basic role in the foundation of physics, from quantum statistical mechanics~\cite{PSW2006,GLTZ2006,R2008,LPSW2009} to the black hole information paradox~\cite{HP2007,SS2008,BF2012,LSHOH2013}.
From the viewpoint of random Hamiltonians, 
random states are an ensemble of ground states of random global Hamiltonians~\cite{M1990},
so that their properties are those typically observed in such {\it global} systems at {\it zero} temperature. It is then natural to ask whether they are still observed in systems with a {\it local} structure at {\it finite} temperature.

In this Letter, we extend a study of the unitarily invariant ensemble of states (equivalently the ensemble of ground states of random global Hamiltonians) to the ensemble of thermal states of random global/local Hamiltonians.
We especially investigate the ensemble of thermal states in comparison with the unitarily invariant ensemble.
To this end, we exploit the concept of a {\it state $t$-design}, an ensemble of states 
simulating, up to the order $t$, statistical moments of random states~\cite{RBSC2004,AE2007},
and investigate whether or not a state $t$-design is approximately achievable in random global/local Hamiltonian systems at finite temperature.
This provides an insight into the validity of the foundation of physics using random states or a state $t$-design when the system respects a local structure and is at finite temperature.
This also has importance in quantum information science since random states have a wide range of applications~\cite{L1997,EWSLC2003,RBSC2004,RRS2005,S2006,DCEL2009}, and their approximate generation is one of the central issues~\cite{EWSLC2003,DLT2002,HL2009,DJ2011,HL2009TPE,BHH2012,CHMPS2013,NM2013,NKM2014,NM2014}.

For an ensemble of thermal states in random {\it global} Hamiltonian systems,
we show that the ensemble monotonically approaches the unitarily invariant one with decreasing temperature and 
that a state $t$-design is approximately achieved at $O(1/{\rm log}(t))$ temperature. 
We then show that, for an ensemble of thermal states in random {\it local} Hamiltonian systems, the ensemble is a state $1$-design at any temperature. We numerically study how close the ensemble is to higher designs and show that the ensemble quickly approaches the unitarily invariant one in a high-temperature regime, but converges to a non-uniform distribution at low temperature. We also give numerical evidence that these two regimes of the ensemble are separated by a singular point, indicating a phase transition of the ensemble at finite temperature.
Since the singularity is not observed for random {\it global} Hamiltonians, this is an intrinsic characteristic of random {\it local} Hamiltonians.

\section{Random states and State $t$-design}
Let $\mathcal{K}$ be a Hilbert space of dimension $D$. 
Random states $\Upsilon$ are an ensemble of pure states uniformly distributed in Hilbert space with respect to the unitarily invariant measure.
Random states play a fundamental role in physics~\cite{PSW2006,GLTZ2006,R2008,LPSW2009,HP2007,SS2008,BF2012,LSHOH2013}, and
are important resource in quantum information processing~\cite{L1997,EWSLC2003,RBSC2004,RRS2005,S2006,DCEL2009},
however, they cannot be efficiently generated.
Hence, an ensemble of states, called an {\it $\epsilon$-approximate state $t$-design} $\Upsilon_{t}^{(\epsilon)}$ has been studied~\cite{EWSLC2003,DLT2002,HL2009,DJ2011,HL2009TPE,BHH2012,CHMPS2013,NM2013,NKM2014,NM2014}.
An $\epsilon$-approximate state $t$-design is defined by $\| \mathbb{E}_{\Psi \in \Upsilon_{t}^{(\epsilon)}}[ \Psi^{\otimes t} ] - \mathbb{E}_{\Psi \in \Upsilon}[ \Psi^{\otimes t} ] \|_1 \leq \epsilon$~\cite{RBSC2004,AE2007}. Here, $\Psi=\ketbra{\Psi}{\Psi}$, $\mathbb{E}$ represents an expectation over an ensemble, i.e. $\mathbb{E}[f(\Psi)]=\int f(\Psi) d\mu(\Psi)$ for the uniform measure $d\mu$, and $\| A \|_1 =\tr |A|$ is the trace norm.
The $\mathbb{E}_{\Psi \in \Upsilon}[ \Psi^{\otimes t}]$
is calculated to be $\Pi_{\rm sym}^{(t)}/d_{\rm sym}^{(t)}$ using Schur's lemma~\cite{GR1999}, where $\Pi_{\rm sym}^{(t)}$ is a projection operator onto a symmetric subspace of $\mathcal{K}^{\otimes t}$ and 
$d_{\rm sym}^{(t)} = \tr \Pi_{\rm sym}^{(t)} = \binom{D+t-1}{t}$.
When $\epsilon=0$, a state $t$-design is called {\it exact} and we denote it by $\Upsilon_t$.
Since a state $t$-design converges to random states when $t \rightarrow \infty$,
the distance between a given ensemble of states and a state $t$-design provides a measure of the uniformity of the ensemble.

\section{Random Global and Local Hamiltonians}
We define random Hamiltonians using the Gaussian unitary ensemble GUE$(L)$, which is an ensemble of $L \times L$ Hermitian matrices $\{ H \}$ 
distributed according to the Gaussian measure $d\mu (H)$ with density proportional to $\exp[-\frac{L}{2} \tr H^2]$~\cite{M1990}. 
We call the GUE the ensemble of {\it random global Hamiltonians} since it has no local structures.
An important feature of random global Hamiltonians is that they are invariant under unitary conjugation, i.e.
$d\mu (u H u^{\dagger}) = d\mu(H)$ for any $u \in \mathcal{U}(L)$ where $\mathcal{U}(L)$ is the unitary group of degree $L$. Hence, their ground states are random states.

We also introduce the ensemble of {\it random $k$-local Hamiltonians}: consider a system consisting of $n$ particles, where the dimension of each particle is $d$. We denote by $\mathcal{H}=(\mathbb{C}^d)^{\otimes n}$ the corresponding Hilbert space. A Hamiltonian $H = \sum_{E} h_E$ is called $k$-local if each term $h_E$ acts nontrivially on a set $E$ of at most $k$ particles.
An ensemble of $k$-local Hamiltonians $\mathfrak{H}_k$ is called {\it random} when each $h_E$ is independently chosen from ${\rm GUE}(d^k)$. Note that $\mathfrak{H}_n$=GUE$(d^n)$ is the ensemble of random global Hamiltonians. Unlike random global Hamiltonians, random $k$-local Hamiltonians for $k\neq n$ do not have global unitary invariance and the ensemble of ground states differs from random states.

At finite temperature $T$, 
a state of a system at thermal equilibrium is given by a thermal state $\rho_H(\beta):=e^{- \beta H}/Z_H(\beta)$, where $\beta =1/T$ is the inverse temperature and $Z_H(\beta)=\tr e^{- \beta H}$ is the partition function.
Although a thermal state is in general not a pure state, we straightforwardly extend the definition of a $t$-design to a mixed state and define a distance between an ensemble of thermal states $\{\rho_H(\beta)\}_{H \in \mathfrak{H}_k}$ and a state $t$-design by
$T_t^{(k)}(\beta) := \frac{1}{2} \| \mathbb{E}_{H \in \mathfrak{H}_k} [ \rho_H(\beta)^{\otimes t} ]-\mathbb{E}_{\Psi \in \Upsilon}[ \Psi^{\otimes t} ] \|_1$.
If an ensemble of thermal states satisfies $T_t^{(k)}(\beta)=\epsilon/2$ for some $\beta$, average properties of the system can be described by an $\epsilon$-approximate state $t$-design up to order $t$.

\section{Random Global Hamiltonian systems}
We first study the ensemble of thermal states for random global Hamiltonians $\mathfrak{H}_n$.
Since an ensemble of their ground states is unitarily invariant, we investigate how the ensemble converges at finite temperature to the unitarily invariant one with decreasing temperature.
When $t=1$, $\mathbb{E}_{H \in \mathfrak{H}_n} [ \rho_H(\beta)^{\otimes t} ]$ reduces to the completely mixed state $I_D/D$, where $D=d^n$ and $I_D$ is the identity matrix in $\mathcal{H}$, since it commutes with all $u \in \mathcal{U}(D)$ due to the unitary invariance of $\mathfrak{H}_n=$GUE($D$).
Hence, $T_1^{(n)}(\beta)=0$ for any $\beta$, implying that the ensemble of thermal states is a state $1$-design at any temperature. For $t \neq 1$, we show below that the distance $T_t^{(n)}(\beta)$ for any $t$ monotonically decreases when $\beta$ increases. 

For simplicity, we denote $\mathbb{E}_{\mathfrak{H}_n}[ (\rho_H (\beta))^{\otimes t}]$ by $X(\beta)$.
Due to the unitary invariance of the GUE and the invariance of the partition function under unitary conjugation, $X(\beta)$ commutes with any unitary matrices of the form of $u^{\otimes t}$ ($u \in \mathcal{U}(D)$).
From Schur-Weyl duality~\cite{GR1999}, $X(\beta) = (\lambda  \Pi_{\rm sym}^{(t)}) \oplus A$, where $\lambda = \frac{1}{d_{\rm sym}^{(t)}}\tr X(\beta) \Pi_{\rm sym}^{(t)} < 1/d_{\rm sym}^{(t)}$, and $A$ is some operator on the space orthogonal to the symmetric subspace. Hence, we obtain 
$\Pi_{\rm sym}^{(t)} X(\beta) \Pi_{\rm sym}^{(t)} = \lambda(\beta)  \Pi_{\rm sym}^{(t)}$.
Recalling that $\mathbb{E}_{\Psi \in \Upsilon^{(t)}}[ \Psi^{\otimes t} ] =  \Pi_{\rm sym}^{(t)}/ d_{\rm sym}^{(t)}$, $T_t^{(n)}(\beta)$ is divided into two terms;
$T_t^{(n)}(\beta) =|\!| (\mathbb{I}^{(t)} - \Pi_{\rm sym}^{(t)}) X(\beta) |\!|_1/2   + |\!| \Pi_{\rm sym}^{(t)} (  X(\beta) - \Pi_{\rm sym}^{(t)}/ d_{\rm sym}^{(t)} ) |\!|_1/2$, where $\mathbb{I}^{(t)}$ is the identity operator on $\mathcal{H}^{\otimes t}$.
Using $\lambda(\beta) \geq  1/d_{\rm sym}^{(t)}$, $T_t^{(n)}(\beta)$ is given by
\begin{equation}
T_t^{(n)}(\beta) = 1-  \tr X(\beta) \Pi_{\rm sym}^{(t)} . \label{Eq:distance}
\end{equation}

We express the projector $\Pi_{\rm sym}^{(t)}$ as 
$\frac{1}{t!} \sum_{\sigma \in S_t} V_{\sigma}$, where $S_t$ is the permutation group of order $t$ and $V_{\sigma}$ is a unitary representation of $\sigma$. Using $\tr V_{\sigma_c} \rho^{\otimes t} = \tr \rho^{|c|}$  for a cyclic element $\sigma_c$ in the permutation group, where $|c|$ is the order of the cycle, $T_t^{(n)}(\beta)$ is rewritten as a function of purities of thermal states as follows;
\begin{equation}
T_t^{(n)}(\beta) = 
1-\frac{1}{t!} \sum_{\sigma \in S_t} \mathbb{E}_{\mathfrak{H}_n}  \prod_{\sigma_c \in \sigma} \tr \bigl(\rho_H (\beta) \bigr)^{|c|}, \label{Eq:cycle}
\end{equation}
where the product is taken over all cycles in $\sigma$.

We finally show that $\frac{\partial}{\partial \beta} T_n^{(t)} (\beta) \leq 0$ for any $\beta$, which implies a monotonic decrease of $T_n^{(t)} (\beta)$ with respect to $\beta$.
It suffices from Eq.~\eqref{Eq:cycle} to show $\frac{\partial}{\partial \beta} \tr \bigl(\rho_H (\beta) \bigr)^{m} \geq 0$ for any natural number $m$. This simply holds since
$
\frac{\partial}{\partial \beta} \tr \bigl(\rho_H (\beta) \bigr)^{m}
= m \frac{Z_H(m\beta)}{\bigl( Z_H (\beta) \bigr)^m} [ \langle H \rangle_{\beta} -  \langle H \rangle_{m \beta}]
$,
where $\langle H \rangle_{\beta}:=\tr [ H \rho_H (\beta)]$ is an internal energy of $H$ at the inverse temperature $\beta$,
and the internal energy satisfies 
$\langle H \rangle_{\beta} \geq  \langle H \rangle_{\beta'}$ for $\beta \leq \beta'$.
The equality holds if and only if $\beta=0, \infty$. 

When $\beta=0$, a thermal state $\rho_H(\beta)$ is the completely mixed state $I_D/D$, so that
$T_t^{(n)}(0)=1-d_{\rm sym}^{(t)}/D^{t}$, which is approximately given by $1-1/t!$ for a constant $t$.
On the other hand, $\lim_{\beta \rightarrow \infty} T_n^{(t)}(\beta)=0$ since ground states of random global Hamiltonians are random states. Hence, $T_t^{(n)}(\beta)$ monotonically decreases from $1-1/t!$ to zero with decreasing temperature.

\begin{figure}[tb!]
\begin{center}
\includegraphics[width=42mm]{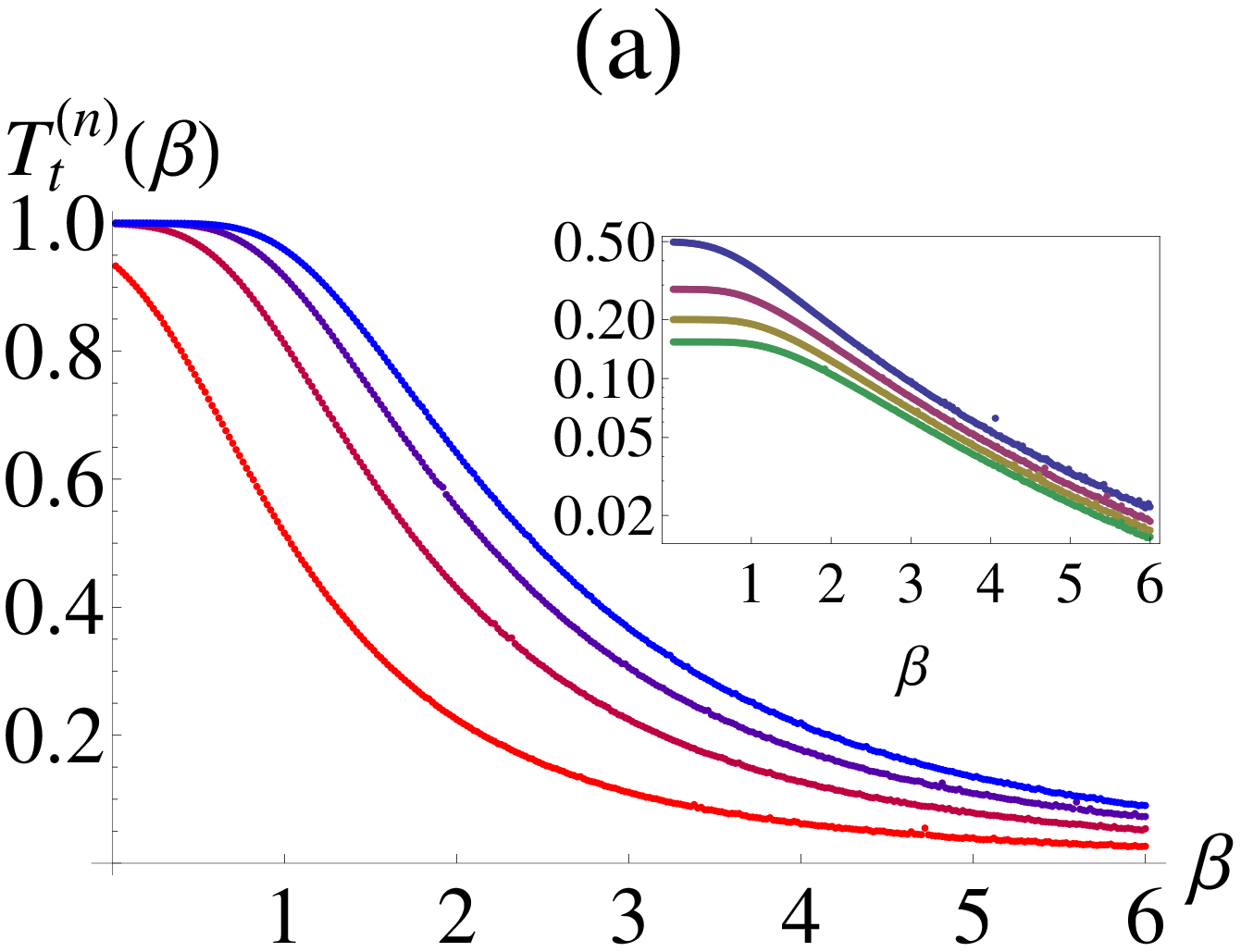}
\includegraphics[width=42mm]{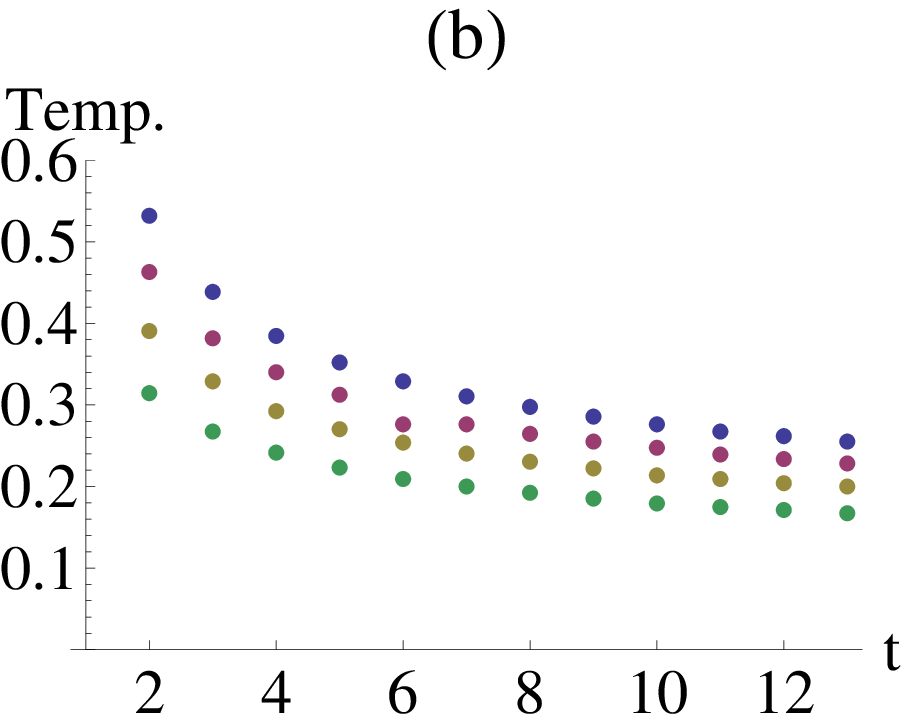}
\caption{
(Color  online) Panel (a) shows upper bounds on $T_t^{(n)}(\beta)$ for $D=4$, where $t=11,8,5,2$ from top to bottom.
The inset shows the scaling of $T_t^{(n)}(\beta)/t$ with $\beta$ for $t=3,6,9,12$ from top to bottom, 
which indicates that $T_t^{(n)}(\beta)$ scales as $t e^{-c \beta}$ for large $\beta$.
Panel (b) shows a sufficient temperature for thermal states of $\mathfrak{H}_n$ to be an $\epsilon$-approximate $t$-design for
$\epsilon=0.5,0.4,0.3,0.2$ from top to bottom. 
}
\label{Fig:UBofDistance}
\end{center}
\end{figure}

For sufficiently large $t$ and $\beta$, $T_t^{(n)}(\beta)$ can be further calculated from Eq.~\eqref{Eq:distance}.
Let $H=\sum_{m=0}^{D-1} E_m \ketbra{E_m}{E_m}$ be an eigendecomposition of $H$, where the eigenvalues satisfy $E_i \leq E_j$, for $i \leq j$.
Using this notation,  Eq.~\eqref{Eq:distance} is rewritten as
$
T_t^{(n)}(\beta) = 1-  \tr \mathbb{E}_{\mathfrak{H}_n}[ \sum \prod_{j=1}^t p_{m_j}(\beta) \bigotimes_{j=1}^t \ketbra{E_{m_j}} {E_{m_j}}  \Pi_{\rm sym}^{(t)} ]
$, where the summation runs over all $m_1, \cdots, m_t \in \{ 0,\cdots D -1\}$.
Since $\tr \bigotimes_{j=1}^t \ketbra{E_{m_j}} {E_{m_j}}  \Pi_{\rm sym}^{(t)} =1$ when $m_i = m_j$ for all $i,j \in \{1,\cdots, t\}$
and is at most $O(1/t)$ otherwise, $T_t^{(n)}(\beta)$ is simply given by 
$T_t^{(n)}(\beta) = 1-  \mathbb{E}_{\mathfrak{H}_n}[ \sum_{m} (p_{m}(\beta))^t] - O(1/t)$, where we have used that $\mathbb{E}_{\mathfrak{H}_n}[ \ketbra{E_{m}} {E_{m}}^{\otimes t}] = \Pi_{\rm sym}^{(t)}/d_{\rm sym}^{(t)}$ for any $m \in \{0,\cdots, D-1\}$.
When $\beta$ is sufficiently large such that $t \beta  \gg 1/\Delta E$ where $\Delta E=E_1-E_0$, $T_t^{(n)}(\beta)$ is approximately given by 
\begin{equation}
T_t^{(n)}(\beta) = 1-  \mathbb{E}_{\mathfrak{H}_n}[(p_{0}(\beta))^t]. ~\label{Eq:EB}
\end{equation}
This provides, in general, an upper bound for $T_t^{(n)}(\beta)$, and it becomes exact when $\beta \rightarrow \infty$. Since  the joint probability distribution of $\{E_k \}_{k=0}^{D-1}$ for $\mathfrak{H}_n$ is known~\cite{HP2007,SS2008,BF2012,LSHOH2013}, an upper bound of $T_t^{(n)}(\beta)$ can be numerically (but exactly) calculated as given in Fig.~\ref{Fig:UBofDistance}.

From Eq.~\eqref{Eq:EB}, we also obtain the scaling of a threshold temperature $T_{\epsilon}$, below which the ensemble of thermal states is an $\epsilon$-approximate $t$-design.
Since $(p_{0}(\beta))^t \sim 1-t e^{-\Delta E \beta}$ for large $t$ and $\beta$, $T_t^{(n)}(\beta) \sim O( t e^{-c \beta})$, where $c$ is a constant. Thus, we obtain $T_{\epsilon}=O((\log t + \log 1/\epsilon)^{-1})$.  The numerics show that this holds even for relatively small $t$ (see the inset of Panel (a) and Panel (b) in Fig.~\ref{Fig:UBofDistance}).
Thus, average properties of random global Hamiltonian systems at temperature $O((\log t + \log 1/\epsilon)^{-1})$ are describable by an $\epsilon$-approximate $t$-design. Since an $\epsilon$-approximate $t$-design has important properties of random states, such as a scrambled feature~\cite{HP2007,SS2008,BF2012,LSHOH2013}, even for small $t$, and can be replaced with random states in many quantum informational tasks using them~\cite{L2009}, so does the ensemble of thermal states in random global Hamiltonian systems at the corresponding temperature. 

\section{Random Local Hamiltonian systems}
For random local Hamiltonians, the investigation of $T_t^{(k)}$ is not simple since the ensemble of local Hamiltonians $\mathfrak{H}_k$ does not have global unitary invariance. However, the ensemble of thermal states of $\mathfrak{H}_k$ for any $k$ is still an exact state $1$-design at any temperature as shown below.

The $\mathfrak{H}_k$ still remains invariant under the conjugation of local unitary operations of the form  $\otimes_{l=1}^n u_l$, where $u_l \in \mathcal{U}(d)$.
Hence, the $\mathbb{E}_{\mathfrak{H}_k} [ \rho_H (\beta)]$ commutes with all local unitary matrices,
implying that it is in the commutant of them;
$\mathbb{E}_{\mathfrak{H}_k} [ \rho_H (\beta)] \in ( \otimes_{l=1}^n \mathcal{U}(d) )'$, where $X'$ is the commutant of an algebra $X$.
Since the commutant of the tensor products is the same as the tensor product of the commutants of each algebra, 
$( \otimes_{l=1}^n \mathcal{U}(d) )' =  \otimes_{l=1}^n ( \mathcal{U}(d) )'$~\cite{RD1975},
$\mathbb{E}_{\mathfrak{H}_k} [ \rho_H (\beta)]$ is in $\otimes_{l=1}^n (\mathcal{U}(d) )'$.
Recalling $(\mathcal{U}(d) )' = \{ I_d \}$ and $\tr \mathbb{E}_{\mathfrak{H}_k} [ \rho_H (\beta)]=1$,  we obtain $\mathbb{E}_{\mathfrak{H}_k} \rho_H (\beta)= I_D/D$,
which implies that $\mathfrak{H}_k$ is an exact $1$-design for any $k$ and for any $\beta$.

\begin{figure}[tb!]
\begin{center}
\includegraphics[width=42mm,clip]{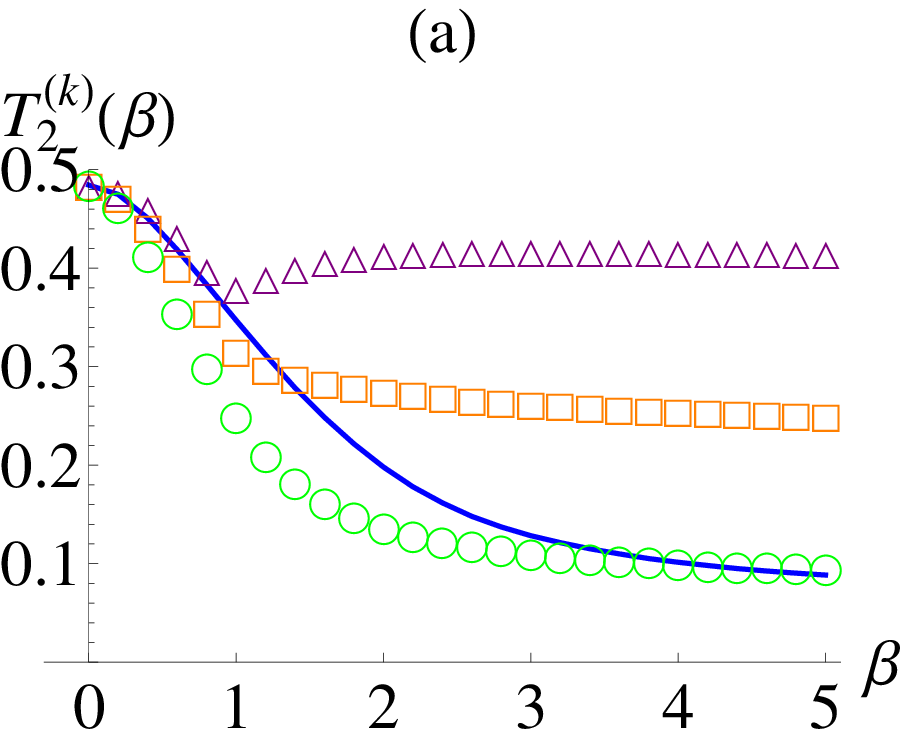}
\includegraphics[width=42mm,clip]{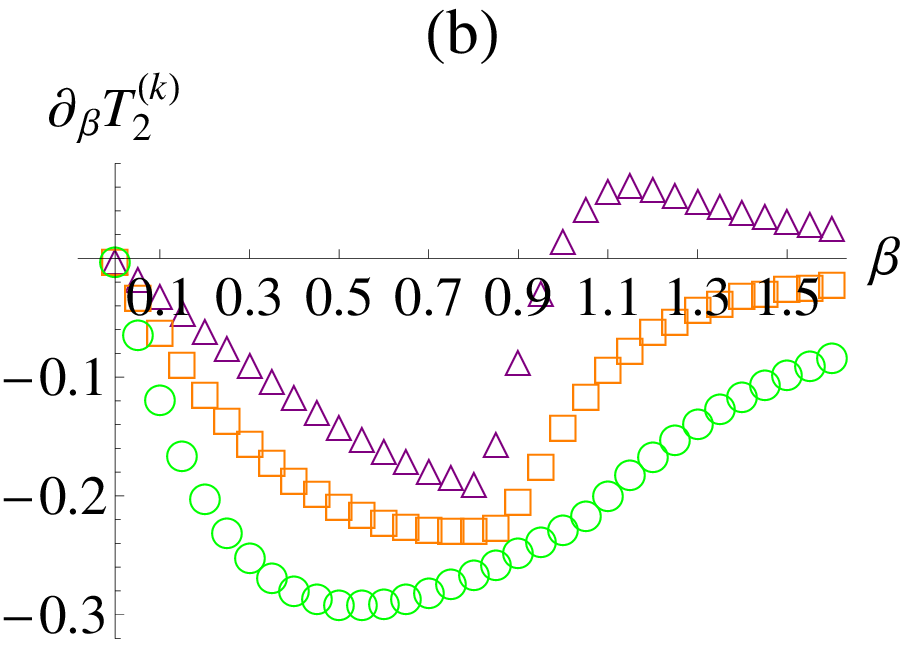}
\caption{
(Color  online) Panel (a) shows the distance $T_t^{(k)}(\beta)$ for $n=5$, $t=2$, and neighboring interactions.
Panel (b) shows its derivative $\partial_{\beta} T_t^{(k)}(\beta)$ in terms of $\beta$.
The purple($\bigtriangleup$), orange($\Box$), green($\bigcirc$) represent $k=2,3,4$, respectively, and the blue solid line is for random global Hamiltonians $\mathfrak{H}_n$.
The expectation of $\mathbb{E}_{\mathfrak{H}_k}[ (\rho_H (\beta))^{\otimes t}]$ is taken by sampling $2\times 10^4$ Hamiltonians. It is observed that $\beta_{\rm c}^{(2)} \sim 0.8$, $\beta_{\rm c}^{(k)} \sim 0.85$, and  $\beta_{\rm c}^{(k)} \sim 1.05$.
}
\label{Fig:localNN}
\end{center}
\end{figure}

We numerically study how close the ensemble of thermal states is to higher designs.
We particularly consider neighboring interactions on a line of qubits, i.e., $d=2$. The results are given for $n=5$ and $t=2$ in Fig.~\ref{Fig:localNN}. It is observed that the distance $T_2^{(k)}(\beta)$ quickly decreases with increasing $\beta$ in a small $\beta$ region. However, when $\beta$ is larger than a certain value, $T_2^{(k)}(\beta)$ is almost constant. This limiting values depend on $k$ and are smaller for larger $k$, which is intuitive since the ensemble becomes random states when $k=n$ and $\beta \rightarrow \infty$.
It is also observed in Fig.~\ref{Fig:localNN} that $T_2^{(k)}(\beta)$ monotonically decreases with $\beta$ when $k \neq 2$. In the case of $k=2$, there exists a dip around $\beta =1$, which is also observed for different $n$.

Panel (b) in Fig.~\ref{Fig:localNN} shows that the two regimes of the ensemble of thermal states, a quickly spreading regime and a converging regime, are likely to be separated by a singular point $\beta_{\rm c}^{(k)}$. This indicates an existence of a phase transition between the two regimes. When $\beta < \beta_{\rm c}^{(k)}$, $\partial_{\beta}  T_t^{(k)}(\beta)$ scales quadratically with $\beta$, while it approaches zero exponentially for $\beta > \beta_{\rm c}^{(k)}$ if $k \neq 2$.  
For $k = 2$, $\partial_{\beta}  T_t^{(k)}(\beta)$ approaches a positive value exponentially and then decreases to zero, reflecting the dip of $T_t^{(2)}(\beta)$.
Although the kink of $\partial_{\beta}  T_t^{(k)}(\beta)$ at $\beta_{\rm c}^{(k)}$ is less prominent for larger $k$, it seems present even for $k=n-1$ but not for $k=n$, where the ensemble converges slowly and smoothly to the unitarily invariant one with decreasing temperature. From these observations, we conjecture that $T_t^{(k)}(\beta)$ ($k\neq n$) has a singular point $\beta_{\rm c}^{(k)}$ in the thermodynamic limit ($n \rightarrow \infty$), leading to a second-order phase transition of the distribution of thermal states.
We also numerically checked that most of the features are present for the case of interactions on a complete graph, except that $T_t^{(k)}(\beta)$ and $\partial_{\beta} T_t^{(k)}(\beta)$ do not monotonically decrease in terms of $k$ in a high-temperature region. 

A possible interpretation of the distinctive temperature $\beta_{\rm c}^{(k)}$, combined with a fact that the expected density of states for the ensemble of random local Hamiltonians is a Gaussian~\cite{HMH2004,KLW2014}, is that the distribution of eigenstates with low energies intrinsically differs from those with intermediate energies.
To explain this clearly, let $P(E) \Delta E \propto \exp[-(\frac{E-\bar{E}}{\sigma})^2] \frac{e^{-\beta E}}{Z(\beta)} \Delta E$ be the population of eigenstates between the energy $[E, E+\Delta E]$ for a small $\Delta E$ in a thermal state, where $\bar{E}$ and $\sigma$ is the mean and the standard deviation of the Gaussian density of states, respectively. 
When $\beta$ is sufficiently small such that the thermal population $e^{-\beta E}/Z(\beta)$ is close to $1/d^n$ for any $E$, the Gaussian term in $P(E)$ is dominant. Hence, the corresponding thermal state is effectively described by a mixture of the eigenstates with eigenenergies in $[\bar{E}- \sigma, \bar{E}+ \sigma]$. On the other hand, the thermal population in $P(E)$ becomes dominant when $\beta$ is large. For instance, the population of eigenstates with energy $E \in [\bar{E}- \sigma, \bar{E}+ \sigma]$ is comparable with that of ground states at $\beta_E = (E-E_0)/\sigma^2$. For $\beta \gg \beta_E$, eigenstates with energy $E$ does not contribute to the thermal state.

Due to this trade off between the Gaussian and the thermal factors in $P(E) \Delta E$, $T_t^{(k)}(\beta)$ in a small (large) $\beta$ reflects the properties of eigenstates with intermediate (low) eigenenergies. 
The $\beta^{(k)}_c$ point is understood as the point where this transition happens. The singularity of $T_t^{(k)}(\beta)$ at $\beta^{(k)}_c$ then indicates that the distribution of intermediate eigenstates for $H \in \mathfrak{H}_k$ is qualitatively different from that of low-energy eigenstates. Since $T_t^{(k)}(\beta)$ rapidly decrease with $\beta$ when $\beta < \beta_{\rm c}^{(k)}$, the distribution of intermediate eigenstates is likely to be similar to the unitarily invariant one.

The situation is entirely different for the ensemble of random global Hamiltonians, where no distinctive temperature is observed, for the following two reasons.
First, there is no trade off between the density of states and the thermal population in $P(E)$ since the density of states obeys the semi-circle law given by $\sqrt{c - (E-E_{\mu})^2}$~\cite{M1990}. This is negligible at any temperature compared to the thermal population that exponentially scales with $E$. Second, the ensemble of eigenstates of any eigenenergy is unitarily invariant, which is in sharp contrast to the distribution of eigenstates of random local Hamiltonians.

These results show that the distribution of thermal states in random $k$-local Hamiltonian systems has a rich structure and is qualitatively different from that in global Hamiltonian systems, even if $k=O(n)$.
This is not only interesting from a theoretical point of view but also physically important since it means that 
most of the previously known results of random states, equivalently an ensemble of ground states in random global Hamiltonian systems, related to physical situations~\cite{PSW2006,GLTZ2006,R2008,LPSW2009,HP2007,SS2008,BF2012,LSHOH2013} cannot be directly applied to many-body systems with local interactions.

\section{Summary and outlook}
In this Letter we have investigated a distribution of thermal states in random global/local Hamiltonian systems.
For random global Hamiltonians, we analytically showed that the ensemble of thermal states monotonically approaches the unitarily invariant one with decreasing temperature and achieves an $\epsilon$-approximate state $t$-design when the temperature is $O(1/(\log t + \log 1/\epsilon))$.
On the other hand, the ensemble of thermal states for random $k$-local Hamiltonians achieves a state $1$-design but not higher designs. 
We then showed by studying a higher design that the ensemble is divided into two regimes of temperature, a regime where the ensemble quickly spreads toward the uniform one with decreasing temperature and a regime where the ensemble converges to a non-uniform one, which are likely to be separated by a singular point.
These studies have revealed the similarities and the differences of random global/local Hamiltonians from the viewpoint of the distribution of thermal states, and have opened a new approach to study random systems by connecting random matrix theory and quantum information science.

 It is desirable to analytically confirm the features numerically observed in this paper. Proving the phase transition of the ensemble of thermal states is especially important. It is also interesting to derive the probability distribution of ground states in random local Hamiltonian systems, by which an understanding of local Hamiltonian systems will be further deepened.

\section{Acknowledgement}
This work was supported by the ERC grants QFTCMPS, and SIQS, and through the DFG by the cluster of excellence EXC 201 Quantum FQ Engineering and Space-Time Research.
Y.N. thanks to R. F. Werner, A. Milsted and C. B\'{e}ny for helpful discussions. 
Y. N. acknowledges JSPS Postdoctoral Fellowships for Research Abroad.

\end{document}